\documentclass[a4paper,fleqn]{cas-dc}

\usepackage[numbers]{natbib}
\usepackage[version=4]{mhchem}
\usepackage{upgreek}
\usepackage{amsmath}
\usepackage{graphicx}
\usepackage{geometry}
\usepackage{hyperref}

\def\tsc#1{\csdef{#1}{\textsc{\lowercase{#1}}\xspace}}
\tsc{WGM}
\tsc{QE}
\tsc{EP}
\tsc{PMS}
\tsc{BEC}
\tsc{DE}

\begin{document}
\let\WriteBookmarks\relax
\def\floatpagepagefraction{1}
\def\textpagefraction{.001}
\shorttitle{The infrared spectrum of dibenzo[a,l]pyrene}
\shortauthors{SD~Wiersma et~al.}

\title [mode = title]{IRMPD spectroscopy of a PAH cation using
FELICE: The infrared spectrum and photodissociation of
dibenzo[a,l]pyrene}                      
\tnotemark[1]

\tnotetext[1]{This document is the results of the research
   project funded by the Dutch Research Council (NWO).}

\author[1,2,3]{SD Wiersma}[%
role=Researcher,
orcid=0000-0002-1793-768X]
\credit{Investigation, Visualization, Writing - Original Draft}

\address[1]{Van `t Hoff Institute for Molecular Sciences, University of Amsterdam, PO Box 94157, 1090 GD, Amsterdam, The Netherlands}
\address[2]{Radboud University, Institute for Molecules and Materials, FELIX Laboratory, Toernooiveld 7, 6525 ED Nijmegen, The Netherlands}
\address[3]{Institut de Recherche en Astrophysique et Planétologie (IRAP), CNRS, Université de Toulouse (UPS), 31028 Toulouse, France}

\author[1]{A Candian}[%
role=Researcher,
orcid=0000-0002-5431-4449]
\credit{Software, Investigation, Writing - review \& editing}

\author[4]{M Rapacioli}[%
role=Researcher,
orcid=0000-0003-2394-6694]
\credit{Software, Investigation, Writing - review \& editing}

\address[4]{Laboratoire de Chimie et Physique Quantiques (LCPQ/IRSAMC), Université de Toulouse (UPS), CNRS, 118 Route de Narbonne, 31062 Toulouse, France}

\author[1]{A Petrignani}[%
role=Co-ordinator,
orcid=0000-0002-6116-5867]
\cormark[1]
\ead{a.petrignani@uva.nl}
\credit{Conceptualization, Resources, Writing - review \& editing, Supervision, Project administration, Funding acquisition}

\cortext[cor1]{Corresponding author}
\begin{abstract}
We present the experimental InfraRed Multiple Photon Dissociation (IRMPD) spectrum and fragmentation mass spectrum of the irregular, cationic PAH dibenzo[a,l]pyrene (\ce{C24H14+}) in the 6--40 $\upmu$m / 250--1650 cm$^{-1}$ range. The use of the the Free-Electron Laser for IntraCavity Experiments (FELICE) enabled us to record its Far-InfraRed (FIR) spectrum for the first time. We aim to understand how irregularity affects the infrared spectrum and fragmentation chemistry of PAHs.  Dibenzo[a,l]pyrene is an asymmetric, non-planar molecule, for which all vibrational modes are in principle IR-active. Calculated harmonic Density Function Theory (DFT) and anharmonic Density Functional based Tight Binding Molecular Dynamics (DFTB-MD)  spectra show a large wealth of bands, which match the experiment well, but with a few differences. The periphery of the molecule contains several edge geometries, but out of all possible modes in the 11--14 $\upmu$m out-of-plane \ce{C-H} bending region, only one band at 13.5 $\upmu$m is prominent. This fact and the richness of the \ce{C-C} stretching range make irregular PAHs a possible contributor to D-class interstellar spectra. The fragmentation mass spectra reveal facile \ce{2H}-loss and no \ce{[2C, 2H]}-loss, which is attributed to the sterically hindered, non-planar cove region, which could protect irregular PAHs from radiation damage.
\end{abstract}




\maketitle

\section{Introduction}
Polycyclic aromatic hydrocarbons (PAHs) are by now commonly accepted as the carriers of the Aromatic Infrared Bands (AIBs), the strongest set of interstellar infrared (IR) emission bands between 3 and 20 $\upmu$m \cite{Tielens2008}. These bands are detected throughout many objects in the interstellar medium (ISM) \cite{Peeters2002,VanDiedenhoven2004,Peeters2004b,Peeters2011}. Astronomical models to analyze and interpret the AIBs are based on mostly theoretical Mid-InfraRed (MIR) spectra of PAHs. However, the spectra of many PAH species are not available or validated. The laboratory spectroscopy of large aromatic species has always been an experimental challenge, specifically for cations. A comprehensive review of methods to study the MIR spectra of cationic PAHs is given in \cite{Oomens2003}. Both infrared multiple-photon dissociation spectroscopy (IRMPD) and Van Der Waals (VDW) dissociation spectroscopy are methods that allow a wide range of PAHs to be studied. Astronomical trends can be derived reliably, as the frequency shifts in the experimental spectra are predictable. The powerful IR free-electron lasers at the FELIX Laboratory (Free-Electron Laser for Infrared eXperiments) in the Netherlands or at CLIO (Centre Laser Infrarouge D’Orsay) in France provide a unique possibility to cover the mid- and far-infrared (FIR) range.  These facilities offer access to Free-Electron Lasers (FELs) via a proposal system open to all, which have made IRMPD and VDW successful and accessible methods to the entire AstroPAH community, \textit{e.g.} \cite{Oomens2000,Lorenz2007,Bakker2011,Jusko2018,Panchagnula2020}. In this work, the unique Free-Electron Laser for IntraCavity Experiments, FELICE, was used to perform IRMPD spectroscopy of a relatively large PAH cation. In addition to the MIR spectrum, the intracavity configuration allows us to probe the FIR spectra and the dissociation mass spectra for undecorated, cationic PAHs. The observed fragmentation behavior is similar to that observed in UV dissociation experiments due to the high excitation levels achieved in IRMPD spectroscopy, making this method a useful proxy for interstellar processes \cite{Wiersma2020,Wiersma2021}.

\begin{figure}[b!]
	\centering
	\includegraphics[width=\linewidth]{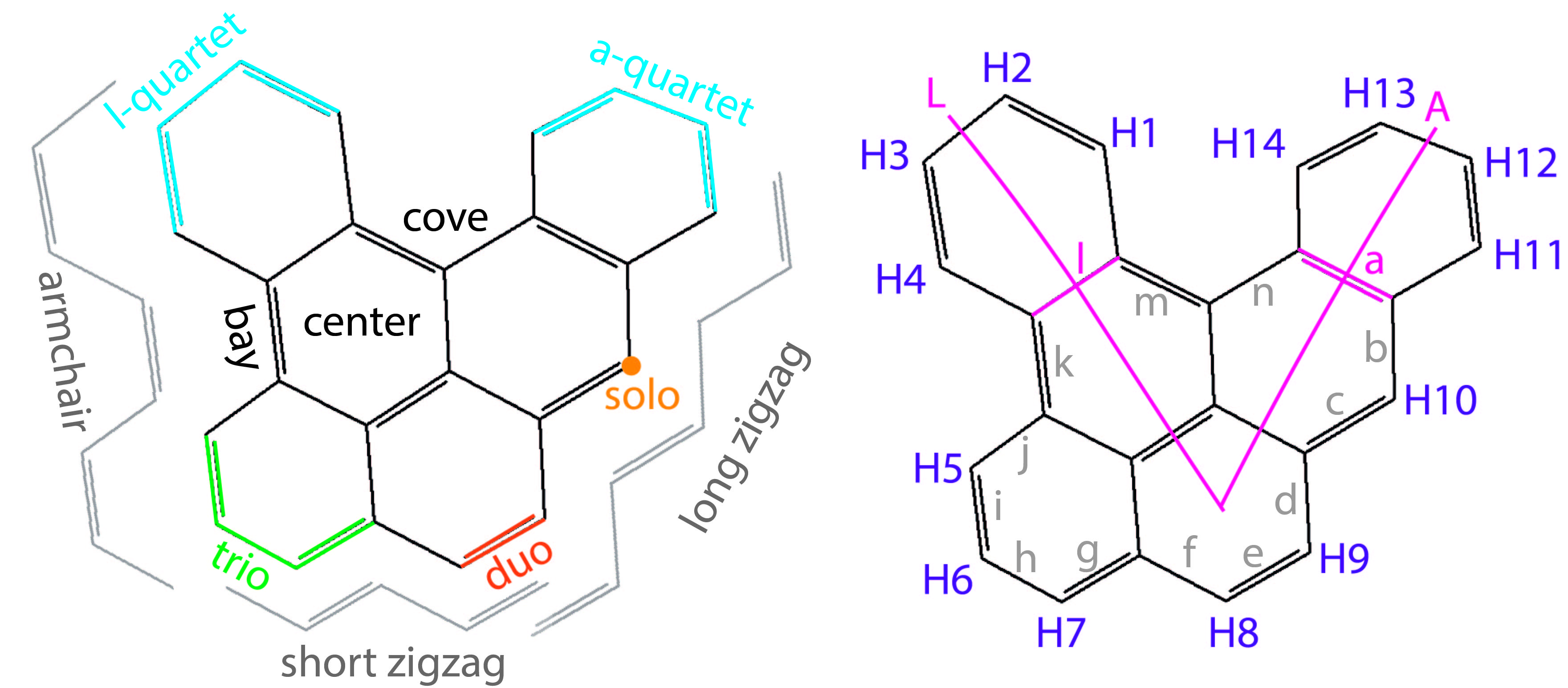}
	\caption{The molecular structure of dibenzo[a,l]pyrene (DBalP), with labeling for the \ce{C-H} and edge peripheries (left), and the labeling for the individual hydrogen atoms, and the pendant [a,l] rings and corresponding vibration propagation axes, based on IUPAC rules (right). }
	\label{fig:numbers}
\end{figure}

Interstellar PAHs occur in various shapes, which are often deduced from the intensity ratios of bands in the 11--14 $\upmu$m range, the spectral region characteristic for the out-of-plane bending modes of aromatic \ce{C-H} bonds (CH$_{\text{oop}}$ modes)  \cite{Hony2001,Hudgins1999,Ricca2010,Ricca2012, Bauschlicher2008,Tielens2008,Bauschlicher2009,Candian2014,Candian2015,Shannon2016}. The left panel in Fig. \ref{fig:numbers} shows these arrangements, and all other periphery types for the molecule studied in this paper, dibenzo[a,l]pyrene (DBalP). The number of neighboring H atoms per ring is the main factor in determining of the peak position in this 11--14 $\upmu$m range, and has led to their designation as being solo, duo, trio or quartet hydrogen atoms \cite{Tielens2008}, based on the spectra of smaller, highly symmetric PAHs \cite{Hudgins1999}. However, the coupling between solo and duo CH$_{\text{oop}}$ modes, and between duo and trio CH$_{\text{oop}}$ modes, can interfere with such distinctions in asymmetric, irregular PAHs \cite{Bauschlicher2008,Bauschlicher2009}. 

For all considered CH$_{\text{oop}}$ bands, the peak positions for cationic PAHs move to shorter wavelengths with respect to their neutral counterparts \cite{Hudgins1999,Bauschlicher2008,Bauschlicher2009,Ricca2012}. 
There is now a rather strong consensus that the strong, astronomically observed 11.2 $\upmu$m band is caused by solo CH$_{\text{oop}}$ vibrations of neutral PAHs, whereas its small 11.0 $\upmu$m side-band is caused by solo CH$_{\text{oop}}$ vibrations of cationic PAHs \cite{Hudgins1999,Hony2001,Bauschlicher2009,Rosenberg2011,Ricca2012,Candian2015}. The origin of the strong 12.7 $\upmu$m-band is less clear. This band is suggested to be of both mixed duo/trio and cationic/neutral character \cite{Hudgins1999,Hony2001,Bauschlicher2008,Shannon2016}, although it was proposed that out-of-plane \ce{C-C-C} (CCC$_{\text{oop}}$) vibrations of armchair-edged, neutral PAHs could also contribute significantly \cite{Candian2014}. PAHs with pendant rings present relatively strong quartet CH$_{\text{oop}}$ modes around 13.5 $\upmu$m \cite{Hudgins1999,Hony2001,Bauschlicher2009,Boersma2010,Rosenberg2011}.  

In spite of the pervasive idea that PAHs with pendant rings are less stable \cite{Gotkis1993,Jochims1994,Jochims1999}, these quartets could be significant contributors to this 13.5 $\upmu$m band \cite{Hony2001}. The postulation that compact, coronene-like PAHs are more stable than irregular ones is mostly based on dissociation studies on small, astronomically unrepresentative PAHs, and should perhaps not be extrapolated. Zigzag edges in room-tempera- ture graphene self-repair to form armchair edges, and armchair edges in PAHs have been proposed to be more resistant to \ce{C2H2}-losses than other edge structures \cite{Chuvilin2010,Poater2007,Koskinen2008,Candian2014,Zhen2014,Zhen2016}. This edge structure also has implications for the hydrogen chemistry inside an astronomical object. Hydrogen atoms in cove and bay regions (see Fig. \ref{fig:numbers} for an example) experience steric hindrance which increases the likelihood of 1,2-hydrogen shifts, and thus forming aliphatic \ce{C-H2} groups, promoting H and \ce{H2} abstraction \cite{Brooks1999,RodriguezCastillo2018,Castellanos2018a}. 

DBalP has many properties that allow us to probe how irregularity can influence the infrared absorption and emission spectrum and how an irregular species fragments. The molecule consists of a cove and a bay region, and one solo, one duo, one trio, and two quartet \ce{C-H} sites, plus an armchair edge and two zigzag edges. It is also non-planar due to the steric hindrance in the cove region. In this paper, we report on a study of the spectral signatures and fragmentation of this highly-irregular, non-planar PAH.
Dibenzopyrene isomers have been investigated in the mid-IR \cite{Bouwman2021}. We expand the spectral coverage of DBalP to the FIR bands, which hold promise to help identifying subclasses of PAHs \cite{Peeters2002,Sloan2014}, and investigate the fragmentation behaviour of this irregular PAHs. 
We recorded the fragment mass spectra and the IR spectrum in the range of 250--1650 cm$^{-1}$, using IRMPD spectroscopy in a Fourier Transform Ion Cyclotron Resonance Mass Spectrometer (FTICR MS) located within the Free-Electron Laser for IntraCavity eXperiments (FELICE).  Additionally, we calculated the harmonic Density Functional Theory (DFT) spectrum and the temperature-dependent anharmonic Density Functional based Tight Binding Molecular Dynamics (DFTB-MD) spectra of DBalP to interpret and compare with the experiment. Finally, we discuss the possible astrochemical impact.
\section{Methods}


\subsection{IRMPD spectroscopy in the FELICE-FTICR MS}
The experiments were performed using the FT-ICR MS beamline of FELICE at the FELIX Laboratory at the Radboud University in Nijmegen \cite{Oepts1995,Militsyn2003,Bakker2010}. A detailed description of this apparatus can be found in \citet{Petrignani2016} and recent upgrades in \citet{Wiersma2021}.

Dibenzo[a,l]pyrene (\ce{C24H14}, $m/z=302$, LGC standards, 99.33\% purity) was brought into the gas-phase at 130 $\textcelsius$C using an effusive sublimation source. The gaseous molecules were ionized with 20 eV electrons from an electron impact ionization source. The resulting ions were directed into a quadrupole mass selector, used in radio-frequency (rf) guiding mode only. They were subsequently led into a quadrupole ion trap with rectilinear rods \cite{Ouyang2004}, which is segmented into three parts for optimal pulse compression. This chamber was pumped down with a turbomolecular pump to $9\times 10^{-7}$ mbar prior to the experiment. In the trap, the ions were collisionally cooled with room-temperature argon gas at a pressure of $\sim10^{-2}$ mbar, for a collection time of 0.05 s.  The thermalized ion cloud was extracted into a large, electrostatic quadrupole bender, which aligns the ion cloud with both the central axis of the FT-ICR and the mirrors of the laser cavity. Through a large, 2 cm radius, 1 m long, rf-guiding quadrupole, the ions were to be stored in the final cell of the four ICR trapping cells, positioned along the central axis. The FELICE waist is widest is cell 4, resulting in the lowest possible, but still sufficient, photon densities, thereby avoiding unnecessary spectral broadening and too stringent alignment constraints. During storage in the ICR cell, masses other than the desired $m/z=302$ were ejected by means of a Stored Waveform Inverse Fourier Transform (SWIFT) pulse \cite{Marshall1998}. FELICE light consists of macropulses of a tuneable length (3--10 $\upmu$s), which in turn consists of a GHz train of ps-length micropulses. The ion cloud was exposed to a single FELICE macropulse, after which the mass spectrum was recorded. The ion intensity of each fragment mass is recorded as a function of the FELICE frequency. Each data point of the recorded mass spectra consists of an accumulation of 20 cycles.
In the 650--1700 cm$^{-1}$ range, the detuning of the undulator was increased from the standard $3\uplambda$ setting to $6\uplambda$, and the macropulse length was decreased from 10 $\upmu$s to 5 $\upmu$s in an effort to reduce saturation.

The IRMPD spectrum was calculated as the ratio of the total fragment ion intensity and the total ion intensity, \textit{i.e.} the sum of both fragment and parent masses. This gives the IRMPD yield $Y(\nu)$ at frequency $\nu$ as:
\begin{equation}
	Y(\nu)=\ln{\left(\frac{N_\text{par}(\nu)}{N_\text{par}\left(\nu\right)+N_\text{frag}\left(\nu\right)}\right)}\text{,}
	\label{eqn}
\end{equation}
for which $N_\text{frag}$ and $N_\text{par}$ are the total fragment and parent mass counts, respectively. The yield $Y(\nu)$ is then divided by $P(\nu)$, the macropulse energy. To this end, a small fraction of the light was coupled out through a 0.5 mm hole in the middle of the cavity end mirror. This light was directed onto a Coherent EPM1000 power meter, and used to measure the laser pulse energy before and after each measurement. A wavelength-dependent factor is then applied for the fluence in the used cell and for the outcoupling of the laser light. Lastly, the resulting power-corrected yields over different wavelength ranges are corrected for their respective baselines, and then normalized to the maximum of the entire curve, resulting in the presented normalized IRMPD yield (Norm. $Y$). Wavelength calibration was performed concurrently with the measurements, by directing the outcoupled beam onto a grating spectrometer (Princeton Instruments SpectraPro). The spectral bandwidth is near transform-limited and $\sim$0.6\% of the full-width at half maximum (FWHM).


\subsection{Computational}
DFT calculations were performed using the Gaussian16 package \cite{g16}. For the harmonic calculations, the B3LYP functional was used \cite{becke,lee}, which has previously been shown to accurately predict the infrared spectra of PAHs \cite{Bauschlicher1997}. A 6-311++G(2d,p) basis set was used to optimize the molecular structure and compute the harmonic vibrational spectrum. A scaling factor of 0.975 was applied to the calculated frequencies to line them up the observed gas-phase spectrum. This factor corrects both for anharmonic redshift and redshift due to temperature and multiple-photon excitation effects. For comparison with the experimental spectrum, the calculated stick spectrum has been convolved with a 20 cm$^{-1}$ Gaussian line shape.

In order to get insights in the effects of temperature on the spectra, we computed the anharmonic spectra for different temperatures from Born-Oppenheimer Molecular Dynamics (MD) simulations. The reader is referred  to a recent publication for methodological details \cite{Chakraborty2021}. Briefly, the anharmonic spectrum can be extracted from an MD simulation in the NVT ensemble (constant number of particles N, volume V and temperature T) from the Fourier Transform (FT) of the dipole autocorrelation function. In order to avoid a pollution of the spectra by the thermostat frequencies, the following strategy is followed.  Several MD simulations are performed in the microcanonical ensemble whose initial conditions are taken from snapshots extracted from an MD  in the canonical ensemble. The spectra are then computed for each of these NVE dynamics (constant number of particles N, volume V and total energy E) and summed up at the end.
In practice, for each temperature, it involved an NVT dynamics of a total duration of 50 ps and 126 NVE dynamics of total duration of 25 ps. A time step of 0.5 fs was used. The choice of the level of theory to compute the potential energy surface results from a compromise between accuracy and computation cost. In this domain the Density Functional based Tight Binding (DFTB) \cite{dftb1,dftb_rev,rev_apx}, an approximated DFT scheme with a much lower computational cost, is our method of choice. The choice of the level of the DFTB scheme and parameters is 
discussed in Appendix \ref{appendixA}.  All DFTB(-MD) calculations have been performed with the deMonNano code \cite{demonnano}.


\section{Results and Discussion}
\subsection{Photodissociation}
\begin{figure}
	\centering
	\caption{The IRMPD mass spectrum of DBalP (\ce{C24H14+}, $m/z = 302$), recorded at 1200 cm$^{-1}$.}
	\label{fig:2bALP-ms}
	\includegraphics[width=0.8\linewidth]{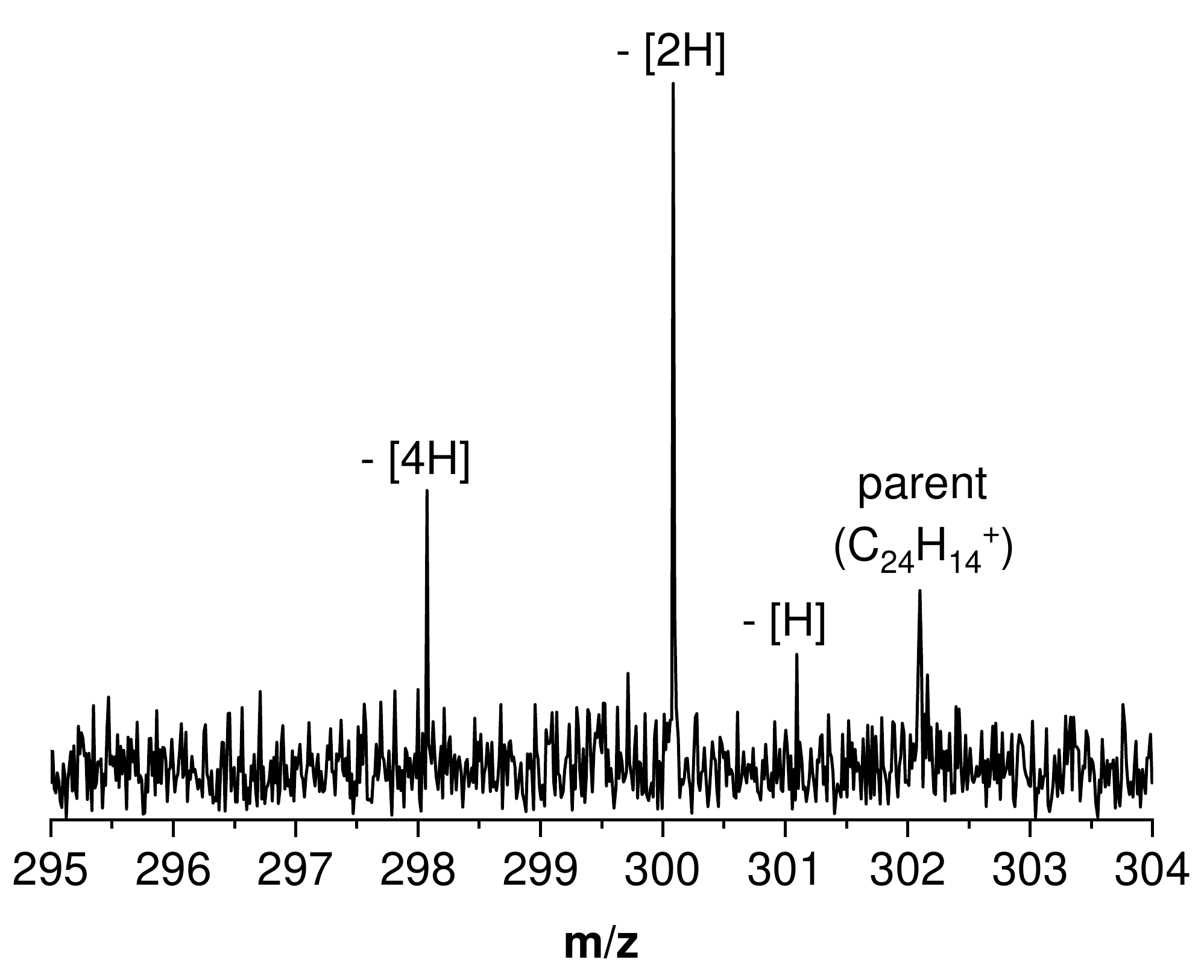}
\end{figure}

Figure \ref{fig:2bALP-ms} shows the IRMPD mass spectrum of cationic DBalP ($m/z = 302$), recorded at 1200 cm$^{-1}$, in the middle of a high-intensity plateau. Next to the nearly-depleted parent ion, three ion fragments are observed, at $m/z = 301$, 300, and 298, corresponding to loss of H, 2H and 4H. At lower-intensity resonances (Norm $Y\lesssim0.05$), small amounts of $m/z = 299$ (--3H) ion signal are observed as well. The loss channel of 2H/\ce{H2} is dominant. No \ce{C2H_x}-fragments are observed.
This is in agreement with previous studies \cite{West2018,RodriguezCastillo2018,Bouwman2021}. These studies also report other isomers of dibenzo[a,l]-pyrene, which do not possess a cove region, to be prone to \ce{C2H_x} fragmentation \cite{RodriguezCastillo2018}. 

Sterically hindered groups in the bay and cove regions of PAHs could play a role in the reaction of H to \ce{H2} in the ISM \citep{Zhen2014,Zhen2017,Zhen2018,Castellanos2018a,RodriguezCastillo2018}.  This hypothesis has been supported through the modeling of interstellar PAHs in photodissociation regions (PDRs) \citep{Boschman2015,Andrews2016}. Furthermore, armchair edges could induce additional structural stability \cite{Candian2014}. This stabilizing effect is, however, not fully understood, as there is currently no explanation for the large \ce{[2C, 2H]}-loss that was observed from cationic isoviolanthrene (IsoV, \ce{C24H18+}) using IRMPD conditions similar to this study \cite{Bouwman2019}. IsoV is an asymmetric molecule with two long armchair edges, but loses much more \ce{[2C, 2H]} than perylene and peropyrene which also contain bay regions an armchair edges, or a compact molecule such as ovalene \cite{Bouwman2019}. The cove region in DBalP is what sets it apart from its two isomers and IsoV. The cove region is likely to be closed into a pentagon after \ce{H2}-loss, making it more resistant to further radiation damage \cite{RodriguezCastillo2018}. We suggest that future photodissociation studies focus on cove regions to find if it can lead to increased stability in larger PAHs as well.  Studies which could verify this pentagon formation would be of particular interest.

\subsection{Infrared spectroscopy}

\begin{figure*}
		\centering
		\caption{The experimental IRMPD spectrum (a) and the scaled harmonic DFT spectrum (b) of DBalP (\ce{C24H14+}). The experimental spectrum has been constructed from two measurement sets, overlapping around 650 cm$^{-1}$. Both (a) and (b) include a magnified inset, displaying the lower-intensity modes in more detail.}
		\label{fig:2bALp}
		\includegraphics[width=0.8\linewidth]{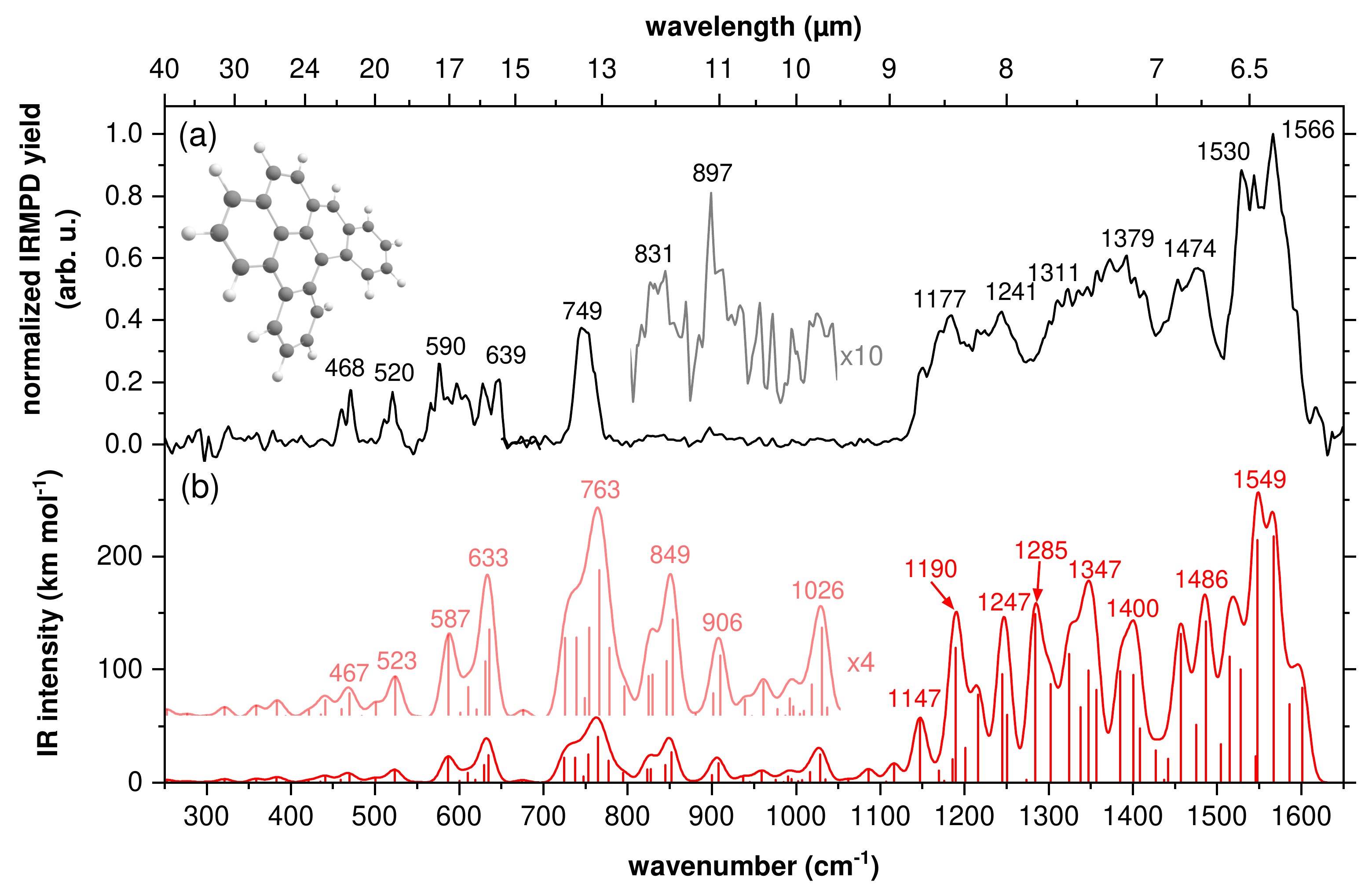}
\end{figure*}

\begin{table*}[width=.9\textwidth,cols=7,pos=h]
	\caption{\label{tab:2bALP-pos} Observed band wavelengths (in $\upmu$m), frequencies (in cm$^{-1}$), band widths (in cm$^{-1}$), and normalized yield $Y$, in comparison with calculated frequencies (in cm$^{-1}$) and IR intensities (in km$\cdot$mol$^{-1}$), with descriptions of the observed modes. Harmonic frequencies are scaled by a factor 0.975. \textit{oop: out-of-plane; ip: in-plane; E/C: elongation/compression; deform.: deformation; antisym.: antisymmetric}}
\begin{tabular*}{\tblwidth}{@{}ccccccl@{}}
		\toprule
		\multicolumn{4}{c}{Experiment} & \multicolumn{2}{c}{Harmonic calculations} & \multicolumn{1}{c}{}                      \\ \cline{1-4} \cline{5-6} 
		Wavel. & Freq.  & Norm. $Y$   & FWHM   & Freq.  & IR int.    & Mode description                      \\
		($\upmu$m)   & (cm$^{-1}$) & (arb. u.) & (cm$^{-1}$) & (cm$^{-1}$) & (km$\cdot$mol$^{-1}$) &                                       \\ \midrule
		21.4   & 468    & 0.12      & 18     & 469    & 7          & ring scissor at short zigzag          \\
		19.2   & 520    & 0.13      & 15     & 523    & 11         & antisym. E/C at short zigzag          \\
		16.9   & 590    & 0.18      & 45     & 587    & 23         & armchair CCC$_{\text{oop}}$/CH$_{\text{oop}}$ all, except H2, 5, 6, 12        \\
		15.6   & 639    & 0.16      & 26     & 635    & 24         & antisym. E/C pendant rings            \\
		13.4   & 749    & 0.38      & 24     & 765    & 41         & CH$_{\text{oop}}$ trio and quartet of L-ring      \\
		&        &           &        & 777    & 19         & CH$_{\text{oop}}$ quartet of A-ring               \\
		12     & 831    & 0.03      &        & 852    & 27         & CH$_{\text{oop}}$ duo, trio, H1, sym./H3, 4, 10 antisym.       \\
		11.1   & 897    & 0.05      &        & 908    & 17         & CH$_{\text{oop}}$ solo/deform. ring A/breathing other rings                 \\
		&        &           &        & 1028   & 25         & antisym. breath. of pendant rings     \\
		&        &           &        & 1147   & 57         & CH$_{\text{ip}}$ scissor duo                      \\
		8.5    & 1177   & 0.36      & 54     & 1190   & 119        & CH$_{\text{ip}}$ rock long zigzag/scissor ring A  \\
		8.1    & 1241   & 0.38      & 59     & 1245   & 96         & breath. central ring                  \\
		&        &           &        & 1284   & 149        & deform. long zigzag and arm.          \\
		7.6    & 1311   & 0.29      & 53     & 1324   & 114        & deform. bay and cove                  \\
		&        &           &        & 1347   & 99         & deform. short zigzag, A ring rock     \\
		7.2    & 1381   & 0.57      & 98     & 1385   & 98         & deform. L axis, L ring rock           \\
		&        &           &        & 1400   & 95         & deform. L axis, duo rock, L ring rock \\
		&        &           &        & 1457   & 132        & deform. along A and L axes            \\
		6.8    & 1475   & 0.51      & 56     & 1487   & 142        & CH$_{\text{ip}}$ rock all/slight ring deform. all \\
		6.5    & 1530   & 0.61      & 25     & 1548   & 214        & deform. ring A                        \\
		6.4    & 1566   & 0.9       & 49     & 1567   & 218        & deform. ring A                        \\ 
		\bottomrule
	\end{tabular*}
\end{table*}

Figure \ref{fig:2bALp}a displays the experimental IR spectrum of cationic DBalP. A plethora of bands is observed, consistent with the non-planar, asymmetric character of the ion, allowing all vibrational modes to have non-zero intensities. This is especially visible in the 1100--1650 cm$^{-1}$ region, where a multitude of transitions gives rise to a broad structure. The measured MIR bands between 1250--1400 cm$^{-1}$ are saturated due to the high power of FELICE and strong absorptions of the molecule in this region. This saturation leads to an apparent underestimation of the normalized intensity in this range as the yield is maximized but still divided by the high laser fluence. The positions of the all experimental bands were determined by deconvolution using Gaussian curves with an average FWHM of 20 cm$^{-1}$, and are given in Fig. \ref{fig:2bALp} and in Table \ref{tab:2bALP-pos}. 
Over the entire experimental range from 250--1650 cm$^{-1}$, at least fourteen bands can be identified. The spectrum shows typical ionic character with active regions largely located between 6--9 $\upmu$m (1100--1600 cm$^{-1}$) and above 12 $\upmu$m (450--800 cm$^{-1}$) with the strongest bands on the blue side. Between 250--450 cm$^{-1}$ and 800--1100 cm$^{-1}$, little to no activity is observed. Two very weak, but significant features were also observed at 831 and 897 cm$^{-1}$, and are shown more clearly in the magnified inset.

The harmonic DFT spectrum, convolved with a 20 cm$^{-1}$ Gaussian waveform and scaled by a factor of 0.975, is displayed in panel (b). It shows a good match with the experimental IR spectrum. The theoretical peak positions given in the Figure designate the positions of the convolved theoretical vibrational maxima. The underlying individual modes are listed in Table \ref{tab:2bALP-pos}. The dense, rich structure in the 1100--1600 cm$^{-1}$ region agrees well with the observed broad structure in experiment. The fitted band frequencies in both theory and experiment match reasonable well, except for around the 1250--1400 cm$^{-1}$ region, where structure is lost in experiment due to saturation. Towards the FIR, in the range of 250--800 cm$^{-1}$, the experimental bands are quite narrow ($\approx 26$ cm$^{-1}$, compared to $\approx 56$ cm$^{-1}$ towards the MIR). This is unlike predicted in theory, where many bands with similar intensities are predicted, leading to relatively broad features.  This myriad of underlying bands shown in the harmonic theory prediction could in reality interact anharmonically, boosting the intensity of one or few remaining modes. This is particularly interesting in view of the search for fingerprints. 

Table \ref{tab:2bALP-pos} also lists the descriptions of the assigned vibrational modes. Assignment of the experimental bands was determined by matching the highest-intensity theoretical modes coinciding with the corresponding observed band. The listed descriptions give the dominant vibrational character of the assigned theoretical modes. Other, lower-intensity vibrational modes contribute to these bands as well, but have been omitted from the assignment for clarity. To aid in the description of the (primary) modes, the different hydrogen positions on the molecule have been unambiguously labeled following IUPAC rules \cite{Ehrenhauser2015}. We identify two principal `axes' consisting of rows of three fused rings, labeled A and L after the [a] and [l] positions on the central pyrene. This labeled structure, the labeling of the sides of the central pyrene and the numbers of the hydrogen atoms are shown in the right panel of Fig. \ref{fig:numbers}. 

In the following, we discuss the listed positions and identified modes in the framework of conventional wavelengths ranges reported for these type of modes. Also listed in Table \ref{tab:2bALP-pos} are the band positions in $\upmu$m as to best compare the assigned modes of the irregular DBalP to those reported in literature for the more symmetric PAH species.
The observed 16.9 $\upmu$m band in this work is assigned to a mode that mixes a typical CCC$_{\text{oop}}$ motion at the armchair edge \cite{Candian2014} and a  CH$_{\text{oop}}$ mode of most H atoms. In rylenes, a CCC$_{\text{oop}}$ appears at 12.7 $\upmu$m \cite{Candian2014}. The long wavelength at which this mode is observed here, indicates that armchair edge modes are not likely to manifest at the same wavelength for different classes of PAHs. The experimental 15.6 $\upmu$m band from this work is best described as an antisymmetric elongation/compression (E/C) of the two pendant rings. E/C modes are more commonly associated with the far-infrared ($>16$ $\upmu$m) \cite{Mattioda2009,Ricca2010}, but the relatively short wavelength at which it is observed here can both be due to the ion's relatively small size and to the coupling of the pendant rings with the central pyrene \cite{Boersma2010}. 
Further spectral differences relative to symmetric species are mostly expected in the CH$_{\text{oop}}$ mode range (11--14 $\upmu$m) due to the irregular periphery. The highest-intensity band observed here is the 13.4 $\upmu$m  band, which falls into the conventional 13.1--13.6 $\upmu$m range for cationic quartet CH$_{\text{oop}}$ modes \cite{Hudgins1999}. The corresponding broad theoretical feature at 13.1 $\upmu$m comprises several different modes, but the dominant one shows quartet character. This demonstrates that even for a highly irregular PAH such as DBalP, the CH$_{\text{oop}}$ bands remain a suitable diagnostic tool. 
As is common for most PAHs \cite{Peeters2002}, the activity between 6--9 $\upmu$m / 1100--1650 cm$^{-1}$ can be largely attributed to \ce{C-C} stretching and in-plane \ce{C-H} bending. Ring deformations tend to follow either the A or the L vibration propagation axis (see Fig. \ref{fig:numbers}) and modes often couple along one of the planes along those axes. However, a few strong CC$_{\text{ip}}$ stretching modes involving both peripheral rings are also present in the theoretical prediction. 

\begin{figure*}
		\centering
		\caption{a) The DFTB-MD simulations of the DBalP IR spectrum at 200, 400 and 800 K, represented by the black, red and blue curves. b) The experimental IRMPD spectrum. c) The harmonic DFT spectrum, scaled by a factor of 0.975 and convolved with a 20 cm$^{-1}$ FWHM Gaussian curve d) The unscaled, harmonic DFTB spectrum, convolved with a 20 cm$^{-1}$ FWHM Gaussian curve. The colored rectangles are guides to compare the experimental bands  with the intensities in the different simulations. }
		\label{fig:allcurves}
		\includegraphics[width=\linewidth]{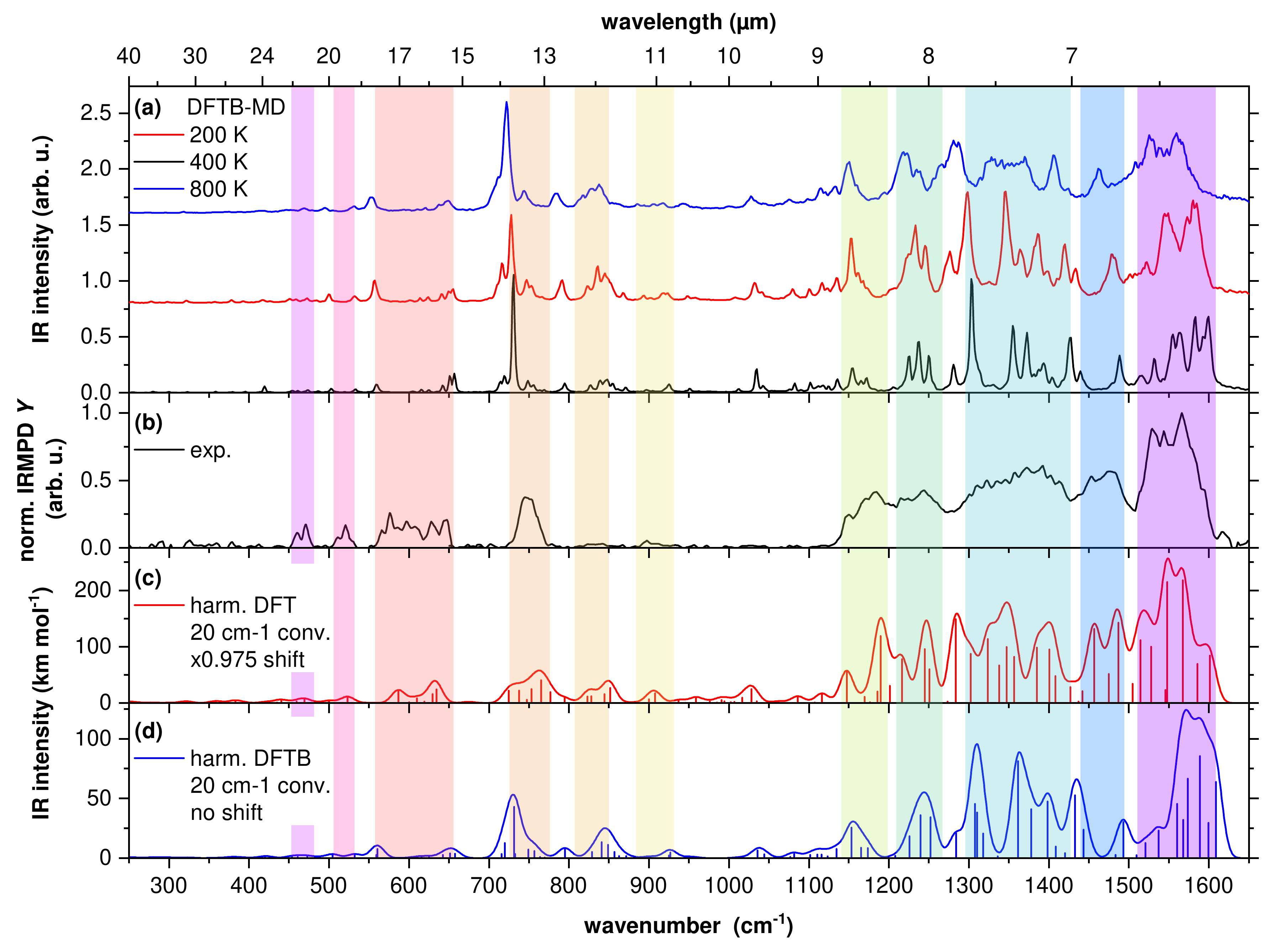}
\end{figure*}

Here, we discuss spectra obtained at the DFTB level. It can be seen from the lower panel of Fig. \ref{fig:allcurves} that the harmonic DFTB spectrum presents the main patterns already 
reported from the experimental and DFT spectra, namely 
a forest of active bands in the MIR region and less intense bands at lower wavenumbers (below 1000 cm$^{-1}$)
as well as an intense band around 730 cm$^{-1}$ also present in the experimental (749 cm$^{-1}$) and DFT spectrum (unscaled harmonic frequency of 784 cm$^{-1}$). As harmonic spectra are stick spectra, they cannot be directly compared to experimental results and a  convolution with a Gaussian or Lorentzian function is usually done. In panels c and d of Fig. \ref{fig:allcurves}, the scaled, harmonic DFT spectrum and the unscaled, harmonic DFTB spectrum are shown, both with a 20 cm$^{-1}$ Gaussian line shape.
The role of the frequency scaling and spectral convolution is, at least in part, to account for anharmonic effects. The values used for the scaling factor and the FWMH are however empirical. As an alternative,  
the thermal evolution of the spectra can be obtained from  DFTB-MD simulations, as  presented in Fig. \ref{fig:allcurves}a and Fig. \ref{fig:DFTB-MD}, where shifts and broadenings are natural, \textit{i.e.} a result of the performed FT, and representative of the physical conditions for heated molecules. 

Focusing first on the dense spectral region between 1000--1700 cm$^{-1}$, it can be seen that, at the lowest temperature (200 K, black), the spectrum consists of a plethora of narrow bands whereas at higher temperatures (400 K, red and 800 K, blue), these bands appear to broaden and merge due to anharmonic effects, resulting in a series of broad plateaus rather than well-resolved structures. This trend is in agreement with the room-temperature experimental spectrum, which also presents similar broad features in this spectral region (see Fig. \ref{fig:allcurves}b). This is only 100 K above the room temperature present in experiment, which is reasonable considering the accuracy of the DFTB-MD simulations. Notably, the double resonance determined in experiment (1530 and 1566 cm$^{-1}$ labels in Fig. \ref{fig:2bALP-ms}), is also present in the DFTB-MD predictions (Fig. \ref{fig:allcurves}).

In the FIR region, the evolution of the spectra with temperature yields only slightly broadened features. The 730 cm$^{-1}$ (749 cm$^{-1}$  in the experiment) band remains pronounced and can still be clearly identified even at the highest simulated temperature of 800 K. This is also in agreement with the experimental spectrum, where a clear contrast  is observed between the dense band population and broadening in the MIR as compared to the FIR.

Another contrast between the FIR and MIR behaviour is that the calculated redshift is relatively small in the FIR as the temperature increases. Redshift is observed throughout the spectrum for all the most identifiable bands, but the 559 cm$^{-1}$ band in the 200 K spectrum shifts by 1.4\% (8 cm$^{-1}$) at 800 K, whereas the 1598 cm$^{-1}$ band shifts by 2.1\% (34 cm$^{-1}$). In the 1100-1700 cm$^{-1}$  region, the strong redshift for the highest energies reduces the MIR active region as the temperature increases. More specifically, the range between the most distant prominent bands shrinks by 35 cm$^{-1}$ from 200 K (1154 and 1599 cm$^{-1}$) to 800 K (1149 and 1559 cm$^{-1}$) Based on the aforementioned considerations, the best match with experiment seems to be the 400 K spectrum. This is relatively close to the room temperature present in experiment.The amount of detail observed at 800 K in the dense 1000--1700 cm$^{-1}$ region is still greater than that observed in the experimental spectrum, allowing us to conclude that saturation and power broadening effects play a large role in the bandwidth, but are not likely to aggravate the redshift.
 
Finally, we would like to point out that the use of a  factor smaller than one to scale the DFT harmonic frequencies as done in Fig. \ref{fig:allcurves}c and Fig. \ref{fig:2bALp}b is consistent with the anharmonic redshift observed for the main MIR bands from DFTB-MD simulations.

\section{Astronomical implications}
\begin{figure}
	\centering
	\caption{IRMPD spectra of cationic DBalP (DBalP; light green) and cationic isoviolanthrene (IsoV; dark green) \cite{Bouwman2019}, and the astronomical IR spectra of A-class object IRAS 23133+6050 (blue) \cite{Sloan2003} and  D-class object IRAS 05110–6616 (violet)\cite{Sloan2003}. The shaded areas denote the \ce{CH_{oop}} vibrations for the solo (yellow), duo (green), trio (blue) and quartet (magenta) groups, as defined for astronomical interpretation by \cite{Hony2001}. The dashed lines denote key features in the D-class spectrum, at 6.2, 7.6, 11.4, 12.0 and 13.1 $\upmu$m. }
	\label{fig:2bALP-compare}
	\includegraphics[width=\linewidth]{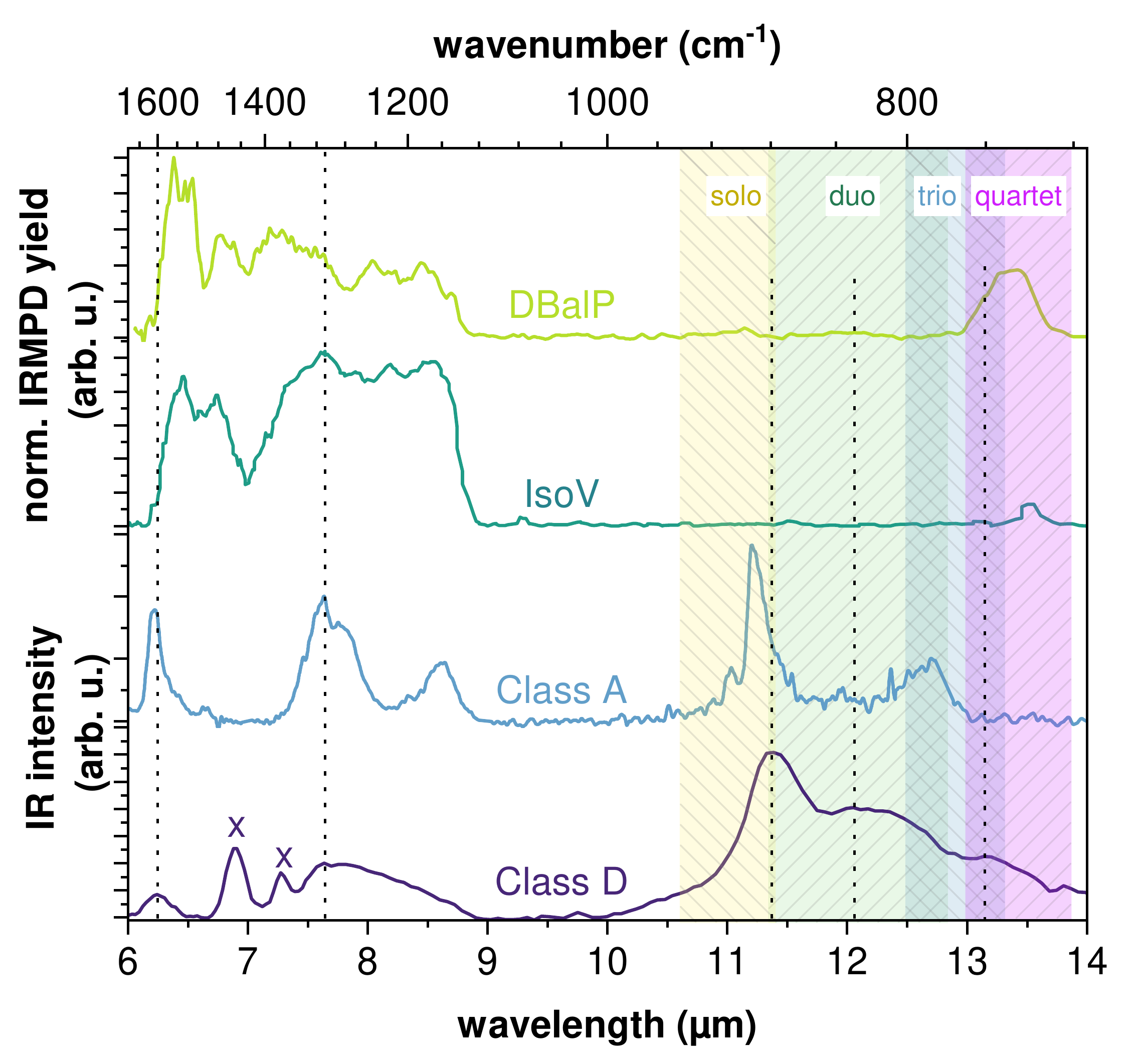}
\end{figure}

Experimental IR spectra of similarly low-symmetry PAHs are scarce \cite{Hudgins1998a,Bauschlicher2009,Maltseva2016,Bouwman2019,Bouwman2021}. To our knowledge there are spectra of three other irregular PAHs in the MIR. In 2019, IsoV was reported, which has a \ce{C_{2h}} symmetry and two long armchair edges \cite{Bouwman2019}. In 2021, the spectra of three dibenzopyrene isomers were reported ([a,e],[a,h] and [a,l]), of which the [a,h] and [a,e] isomers do not have a cove region, but do have armchair edges and \ce{C_{2h}} and \ce{C_{s}} symmetries respectively \cite{Bouwman2021}. 

In Fig. \ref{fig:2bALP-compare}, our DBalP spectrum ([a,l]; light green) is compared to the previously reported IsoV spectrum (dark green) in the 6--14 $\upmu$m range.  The spectral activity of both IsoV and DBalP can be roughly divided in three active regions; 6--7, 7--9, and around 13.5 $\upmu$m. For both cations, the activity between 6--7 $\upmu$m exhibits a double-band structure, albeit with a differing intensity ratios. Furthermore, the activity in 7--9 $\upmu$m is so dense that it approaches a plateau-like feature. The activity around 13.4/13.5 $\upmu$m shows well-defined peaks, although the IsoV feature is markedly sharper. 

\citet{Bouwman2019} also compared the IsoV spectrum to the spectrum of a D-class astronomical object \cite{Peeters2002,Sloan2003,Matsuura2014}. They showed that D-class spectra do not line up with the spectra of more regular PAHs, while irregular PAHs show more similarity in certain aspects. This explanation is also consistent with the hypothesis that PAHs become increasingly photoprocessed in transition from C- and D-class to A- and B-class objects \cite{Sloan2014}. They further argued that because irregular PAHs are underrepresented in the NASA Ames PAH database, fits of astronomical spectra using this database will not be able to accurately represent broad features such as those found in D-class spectra. They concluded that (experimental) spectra of more irregular PAHs are required to make such a fit possible. As an answer to that call, we add the IRMPD spectrum of DBalP to this comparison. 

The same D-class spectrum is shown in Fig. \ref{fig:2bALP-compare} in addition to an A-class spectrum. The distinct astronomical features at 6.9 and 7.3 $\upmu$m are due to aliphatic content that is likely present in this D-class object , and are marked with "x". The vertical dotted lines are placed to guide the eye to the suspected PAH peak positions, at 6.2, 7.6, 11.4, 12.0 and 13.1 $\upmu$m \cite{Sloan2014}. In the 10--14 $\upmu$m range, the solo, duo, trio and quartet \ce{CH_{oop}} vibrational ranges are marked by the shaded areas \cite{Hony2001}. Upon comparing the D-class spectrum to the IsoV and DBalP spectra, a match or significant contribution is not immediately evident. Note that the large 11.3 $\upmu$m feature in the D-class spectrum is associated with solo \ce{CH_{oop}} vibrations, and the broad 12.0--12.7 $\upmu$m feature is likely caused by duo/trio \ce{CH_{oop}} vibrations, neither of which appear in the IsoV or DBalP spectra. However, the 13.1 $\upmu$m feature is likely caused by quartet \ce{CH_{oop}} vibrations from irregular PAHs. 
Furthermore, the long, reddening tail from 7.6 $\upmu$m to 9.0 $\upmu$m visible in the D-class spectrum could contain contributions from the rich spectral range shown for the two irregular PAHs here. Both irregular PAHs show a more pronounced mismatch with the A-class spectrum than with the D-class spectrum. The lowered symmetry leads to a rich spectrum in the 6.1--7.0 $\upmu$m range, which is irreconcilable with (D-class) astronomical spectra, although more MIR spectroscopy on irregular PAHs is needed to confirm that this is true for all irregular PAHs.  

\section{Conclusions and Outlook}
Thanks to the unrivaled power of FELICE, we were able to measure the IRMPD spectrum and mass spectrum of DBalP, all the way down to 250 cm$^{-1}$. 

DBalP's fragmentation mass spectrum reveals facile 2H/\ce{H2}-loss, and no \ce{[2C, 2H]}-loss. This is contrast to the IsoV molecule and other DBalP isomers, which easily fragment through the loss of \ce{C2H_x} under similar conditions. Considering that the latter molecules have armchair edges, we conclude that the presence of armchair edges alone is not sufficient to predict structural stability. Instead, a cove region is required to protect carbon loss, most likely via five-membered ring-formation after hydrogen loss. Currently, DBalP is the only molecule with a cove region of which photofragmentation data is available, so studies on other PAHs with cove regions are of great interest.

We present the IR spectrum of cationic DBalP, and compared it to scaled, harmonic DFT calculations. The spectrum shows 1) a rich \ce{C-C} stretching and CH$_{\text{ip}}$ bending structure in the 1100--1650 cm$^{-1}$ region 2) a CH$_{\text{oop}}$ bending region (700--950 cm$^{-1}$) that is dominated by a relatively strong quartet mode at 749 cm$^{-1}$, and 3) a FIR range with vibrational mode types that are strongly shifted in band position compared to regular, symmetric species.
Overall, a good match between theory and experiment is found in terms of peak position (excluding the saturated range, the average deviation is 10 cm$^{-1}$), but less so in terms of band shape and relative intensity. 

The thermal evolution of the spectrum is simulated with DFTB-MD. At 200 K, it starts with a plethora of bands in the 1100--1650 cm$^{-1}$ region, to evolve into a set of broad, merged bands which have shifted up to 34 cm$^{-1}$ to the red at 800 K. The simulated spectrum at 400 K best reproduces the experimental band width in this highly active region.
Below 1100 cm$^{-1}$, the bands remain distinct and the redshift limited. In both ranges, these trends mirror what is observed for our room-temperature experimental spectrum.

Comparison to a previously reported spectrum of an asymmetric PAH, IsoV, reveals a resemblance in the rich \ce{CC_{ip}} stretching and CH$_{\text{ip}}$ bending structure of the 1050--1650 cm$^{-1}$ region, as well as a typical \ce{CH_{oop}} bending quartet mode at 740/749 cm$^{-1}$. We confirm that DBalP is an unlikely contributor to the spectrum of A-class objects \cite{Bouwman2021}, and are cautious in ascribing the properties of D-class spectra to irregularity as was done before \cite{Bouwman2019}. We put forward that more experimental and precise theoretical research on the spectra of irregular PAHs is needed in order to better understand their astrochemical importance.

\appendix
\section{Computational procedure \label{appendixA}}
\subsection{Choice of the DFTB scheme and parameters}
\begin{figure*}
	\centering
	\caption{Harmonic spectra obtained at the DFT and DFTB levels with various combination of parameter sets (see text for further details) }
	\label{fig:param}
	\includegraphics[width=0.8\linewidth]{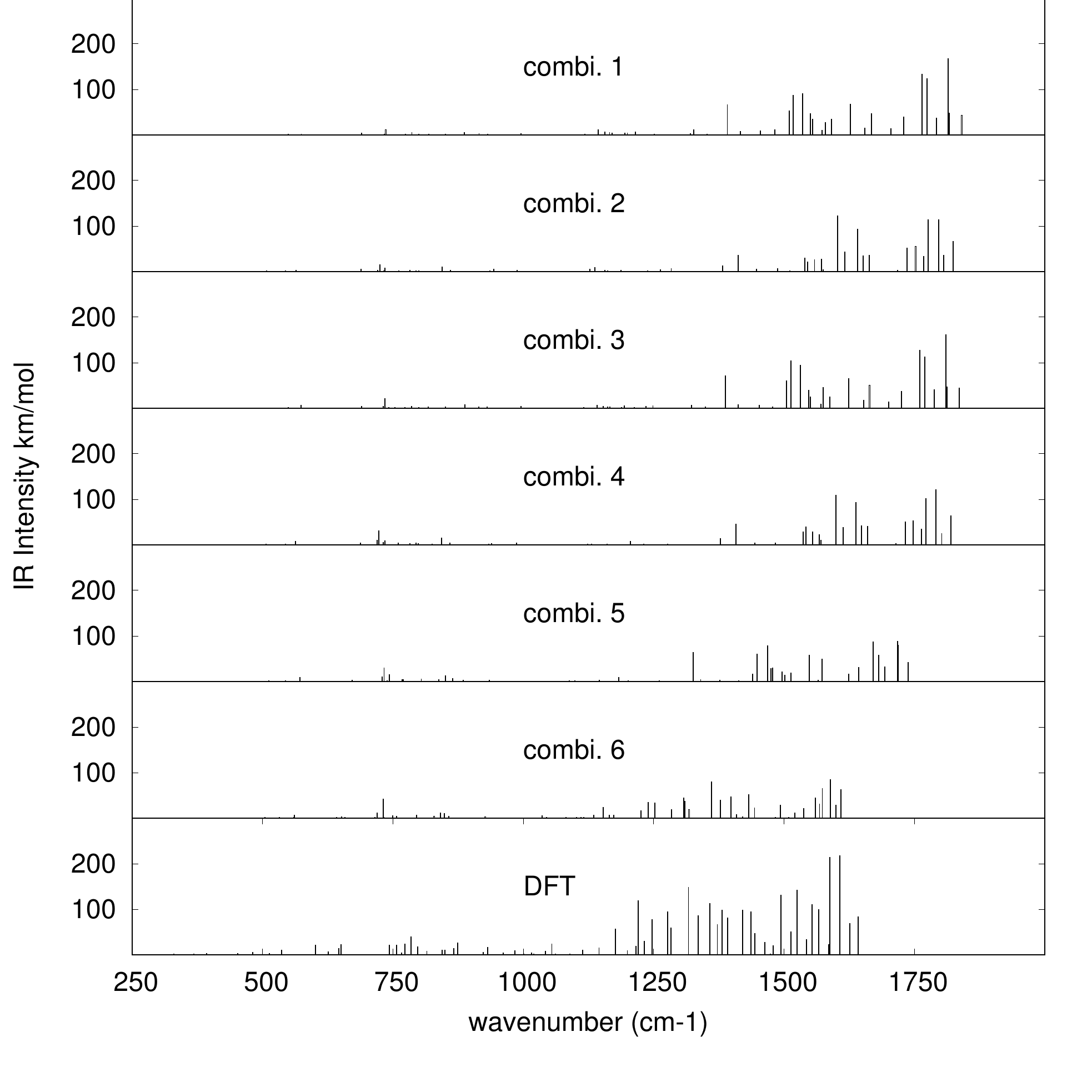}
\end{figure*}

\begin{figure}
	\centering
	\caption{IR spectra of DBalP (C$_{24}$H$_{14}^+$)  for various temperatures from DFTB-MD simulations.   }
	\label{fig:DFTB-MD}
	\includegraphics[width=1.\linewidth]{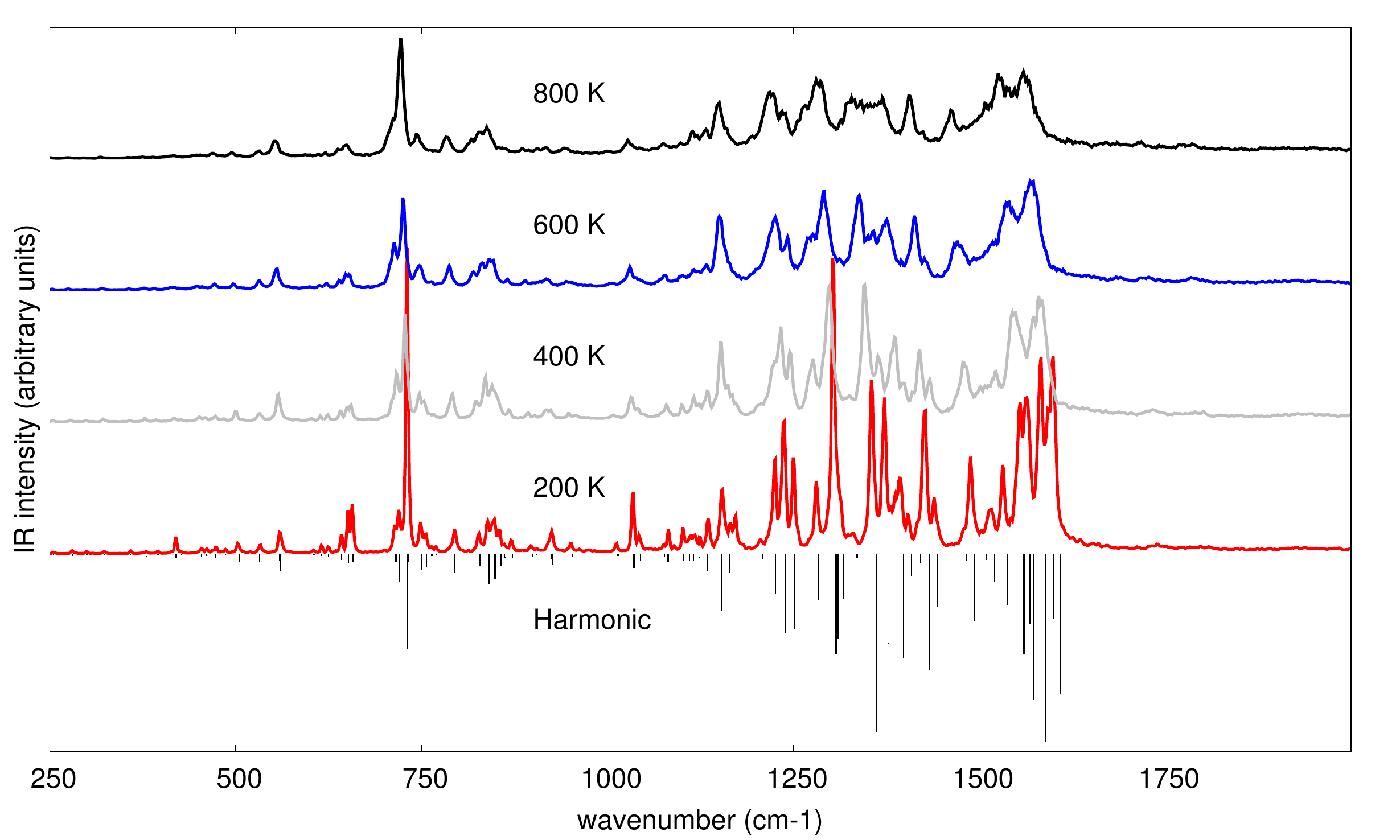}
\end{figure}
\begin{figure}
	\centering
	\caption{Convergency with the number of DFTB-MD simulations of the IR spectra DBalP (C$_{24}$H$_{14}^+$)  for various temperatures. The red (respectively blue) curves are computed from half (respectively the other half) of the performed MD simulations.}
	\label{fig:MDconv}
	\includegraphics[width=1.\linewidth]{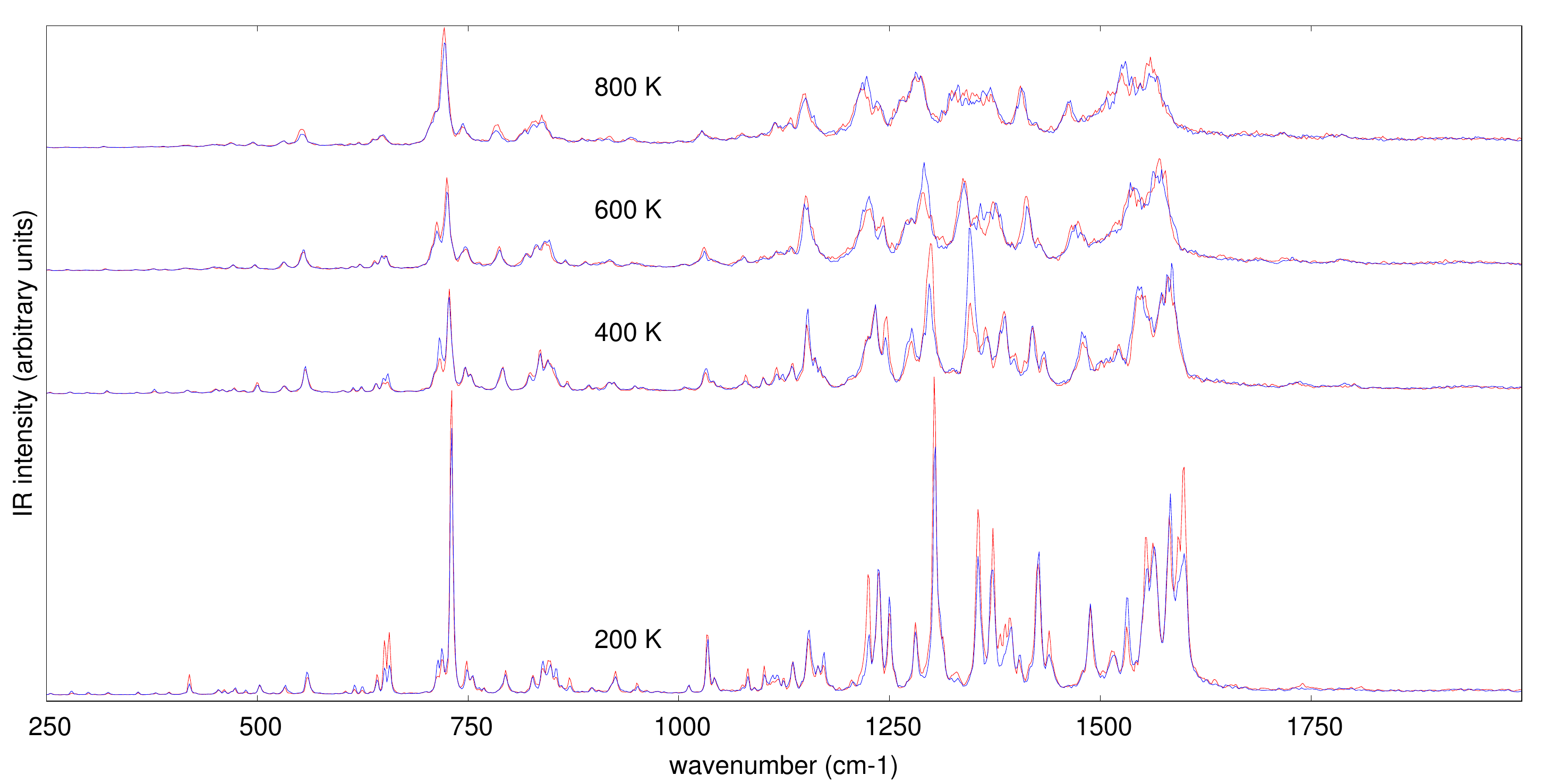}
\end{figure}

The ability of DFTB to reproduce IR spectra  has been addressed in several publications \cite{WITEK:2005aa,Kruger:2005fv,Witek:2004aa}. In particular, it was shown to provide anharmonic factors in agreement with experimental measurements for pyrene and coronene \cite{Joalland2010}. Nevertheless, when addressing a new system, the first step consists
in choosing the appropriate level of  description and parameters. This has been done on the basis of comparison with DFT at the harmonic level, the idea being to have the best match with DFT before extracting anharmonic effects from the MD simulations. Several combinations (comb.)  have been tried:  the self-consistent charge SCC-DFTB level \cite{scc-dftb} using pbc \cite{matsci}  (comb. 1 and 3) or mio \cite{scc-dftb} (comb. 2 and 4) parameters with (comb. 1 and 2) and without (comb 3 and 4) atomic charge corrections \cite{DFTB_CM3}. We also tried the third order DFTB level with a hydrogen bond correction \cite{DFTB3_JPCA2007,Gaus:2011tw} with standard 3ob  (comb. 5) and frequency corrected 3obfreq (comb. 6) parameters \cite{Gaus:2013aa}. The harmonic spectra computed with these combinations are presented in Figure \ref{fig:param} and compared with the harmonic DFT calculations (no scaling factor used in this Figure). 
A forest of active bands can be observed in the MIR region and less intense bands at lower wavenumbers (below 1000 cm$^{-1}$). In addition, a band around 730 cm$^{-1}$ is also present in the DFT spectrum (unscaled frequency of 784 cm$^{-1}$). At the DFTB level, it appears that the positions of the different features strongly depend on the choice of the parameters and  that the combination 6 (third order DFTB with 3obfreq parameters) provides the best compromise in particular concerning the MIR dense region. This combination has been retained to perform the MD simulations used to extract anharmonic spectra.

The convergency of the anharmonic spectra with the number of simulations has also been benchmarked. To do so, the total number of simulations has been shared in two groups and the anharmonic spectra have been computed for each group of simulations. It can be seen in Figure \ref{fig:MDconv} that, although the maximum intensity of the most intense band still presents some fluctuations between the two groups, the trends associated to thermal effects discussed in this paper (shifts, broadening and merging) can be already derived from the red or blue spectra, i.e. using only half of the total number of simulations.
\section{Gaussian fits of the experimental spectrum}
\begin{figure}
	\centering
	\caption{The IRMPD spectrum of DBalP (black) shown together with the individual Gaussian curves (red) which make up the composite fit (blue).}
	\label{fig:fits}
	\includegraphics[width=1.\linewidth]{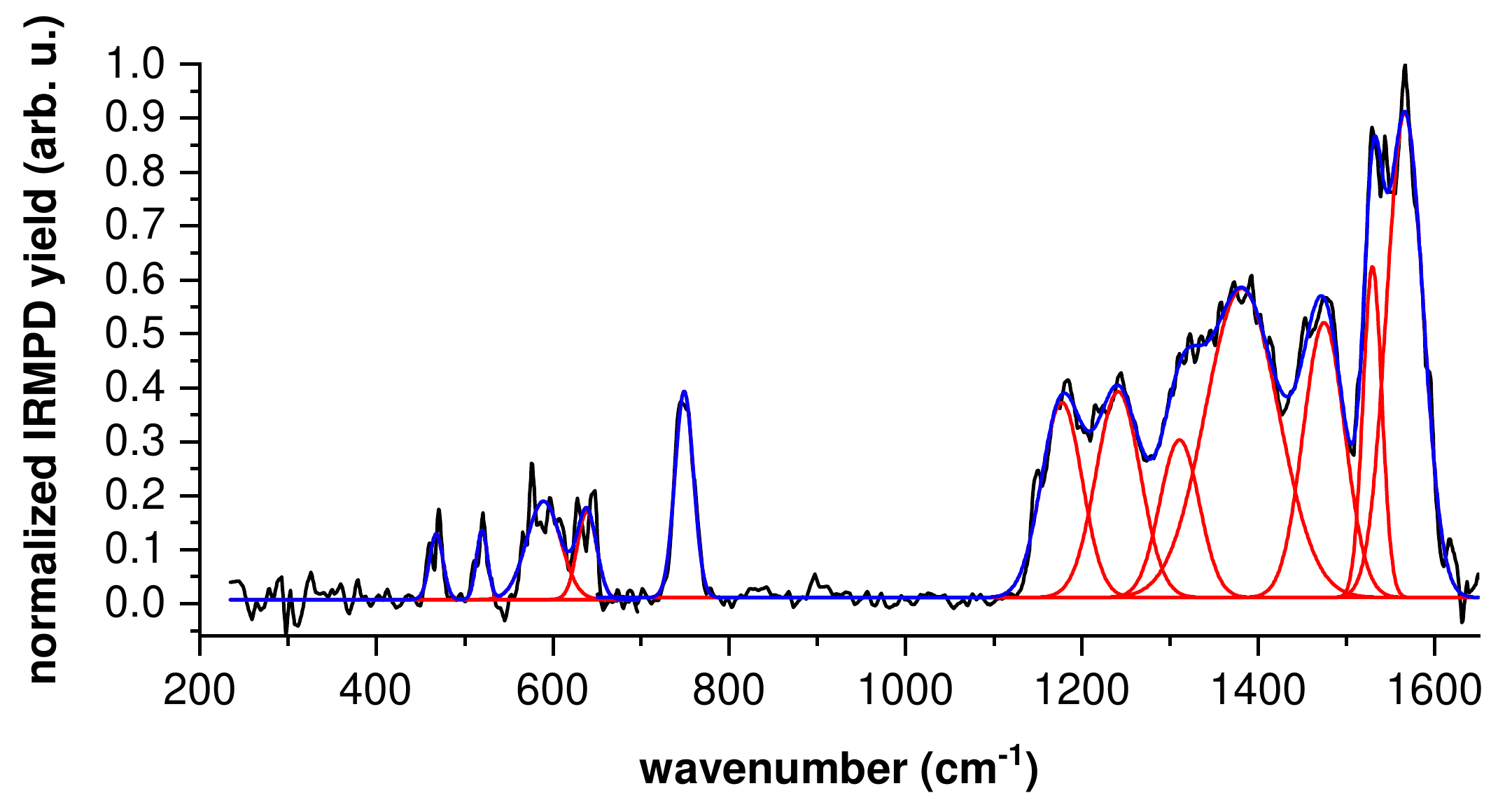}
\end{figure}

The peak positions in the experimental curve were determined through fitting with Gaussian functions of an arbitrary width in cm$^{-1}$, to best approximate the fact that bands with strongly varying intensities are responsible for the shape of this spectrum. The result of this fit is depicted in Fig. \ref{fig:fits}.

\section*{Acknowledgements}
This manuscript was based on a thesis chapter which has been thoroughly read and redacted by prof. Wybren Jan Buma, for which we are very grateful. We also thank dr. J. Bouwman for the helpful discussions on the astronomical context and for sharing his data. We thank COST Action CA18212 - Molecular Dynamics in the GAS phase (MD-GAS), supported by COST (European Cooperation in Science and Technology). We gratefully acknowledge the {\it Nederlandse Organisatie voor Wetenschappelijk Onderzoek} (NWO) for their support of the FELIX Laboratory. This work is supported by the VIDI grant (723.014.007) of A.P. from NWO. Furthermore, A.C. gratefully acknowledges NWO for a VENI grant (639.041.543). Calculations were carried out on the Dutch national e-infrastructure (Cartesius and LISA) with the support of Surfsara, under projects NWO Rekentijd 16260 and 17603
and on the CALMIP supercomputing center (Grant p18009 and p0059).
\printcredits

\bibliographystyle{cas-model2-names}

\bibliography{cas-refs}





\end{document}